\documentclass[aps,superscriptaddress,nofootinbib,twocolumn]{revtex4-1}

\usepackage{mathrsfs}
\usepackage{amsfonts}
\usepackage{amssymb}
\usepackage{amsmath}
\usepackage{graphicx}
\usepackage[usenames,dvipsnames]{color}
\usepackage[colorlinks=true,citecolor=blue,linkcolor=magenta]{hyperref}
\usepackage{ulem}
\usepackage{lmodern}

\newcommand{\figpath}{./}

\newcommand{\ket}[1]{\vert{ #1 }\rangle}

\newcommand{\ceil}{\mathrm{ceil}}
\newcommand{\pro}[1]{p_\text{#1}}
\newcommand{\eps}[1]{\epsilon_\text{#1}}

\begin{document}

\title{Resource costs for fault-tolerant linear optical quantum computing}

\author{Ying Li}

\affiliation{Department of Materials, University of Oxford, Parks Road, Oxford OX1 3PH, United Kingdom}

\author{Peter C. Humphreys}

\affiliation{Clarendon Laboratory, Department of Physics, University of Oxford, Parks Road, Oxford OX1 3PU, United Kingdom}

\author{Gabriel J. Mendoza}

\affiliation{Centre for Quantum Photonics, H. H. Wills Physics Laboratory \& Department of Electrical and Electronic Engineering, University of Bristol, Merchant Venturers Building, Woodland Road, Bristol BS8 1UB, United Kingdom}

\author{Simon C. Benjamin}

\affiliation{Department of Materials, University of Oxford, Parks Road, Oxford OX1 3PH, United Kingdom}

\date{\today}

\begin{abstract}
Linear optical quantum computing (LOQC) seems attractively simple: information is borne entirely by light and processed by components such as beam splitters, phase shifters and detectors. However this very simplicity leads to limitations, such as the lack of deterministic entangling operations, which are compensated for by using substantial hardware overheads. Here we quantify the resource costs for full scale LOQC by proposing a specific protocol based on the surface code. With the caveat that our protocol can be further optimised, we report that the required number of physical components is at least five orders of magnitude greater than in comparable matter-based systems. Moreover the resource requirements grow higher if the per-component photon loss rate is worse than one in a thousand, or the per-component noise rate is worse than $10^{-5}$. We identify the performance of switches in the network as the single most influential factor influencing resource scaling. 
\end{abstract}

\maketitle

\section{Introduction}

Numerous different physical systems have been explored as platforms for quantum information processing. Most approaches involve embodying information in matter systems such as ions or superconducting qubits, but a striking alternative is linear optical quantum computing (LOQC) where all information is encoded in electromagnetic field modes, and processing is carried out using only linear optical elements~\cite{Ralph2010209}. Using light as the information medium takes advantage of the low decoherence suffered by optical fields, and the relative ease with which quantum information can be encoded photonically. However there are drawbacks, in particular the impossibility of deterministic entanglement and the impact of photon loss (whether due to absorption, leakage or detector failure). Such difficulties can be solved by increasing the physical complexity of the circuitry. Thus while LOQC may benefit from simple building blocks, conversely it may require more complex circuits than other approaches, and the balance of these factors will determine whether the approach is a practical competitor to matter-based processors. 

The most developed method for LOQC to date is based on a discrete dual-rail encoding, in which each qubit is encoded in the field modes occupied by a single photon~\cite{Kok2007} (these modes can be spatial, polarisation, time-frequency or any other degree of freedom supported by electromagnetic fields). Crucially, even though entangling operations between dual-rail encoded photonic qubits cannot succeed deterministically, it has been shown that it is nonetheless possible to build an essentially deterministic universal quantum computer using only linear optics. This can be achieved by attempting probabilistic entangling operations (PEO) between many resource states in order to ensure that, with high probability, a sufficient number of operations will succeed to allow for quantum computing~\cite{Nielsen2004,Browne2005}. 

Techniques for mitigating photon loss have also been developed. It has been shown that if quantum information is suitably encoded in a multi-photon state, then losses of up to $50\%$ of the photons can be tolerated before the encoded information is lost~\cite{Varnava2006}. However, in any realistic implementation of a quantum computer, one must account for how complex multi-photon states can be created given that every component, at every level, will be associated with finite rates of photon loss and other forms of noise. Furthermore, the circuitry associated with overcoming non-deterministic entanglement will require many linear optical elements, including delay lines and switching networks, in order to dynamically reroute the outputs of successful operations to the next stage of processing. These elements will induce further errors and losses, and in this sense the twin issues of non-deterministic entanglement and photon loss aggravate one another in LOQC. Fortunately the threshold theorem assures us that, if all physical error rates are sufficiently low then errors at the logical level can be made arbitrarily rare, and scalable fault-tolerant quantum computing can be achieved. The central challenge of quantum computing is therefore to demonstrate the operations necessary for quantum computing with error rates below these thresholds. Theoretical studies have established the required thresholds for architectures relevant to superconducting qubits~\cite{fowlerSCarch}, and to matter optical networks~\cite{BenjPRX}, and experimental systems have been demonstrated at, or beyond, the required performance levels~\cite{Lucas1,Lucas2,Martinis2015}. However to our knowledge no prior paper has established requirements of LOQC at the per-component level while simultaneously tracking the overall resource costs. 

\begin{figure*}[!]
\centering
\includegraphics[width=1.0\linewidth]{\figpath 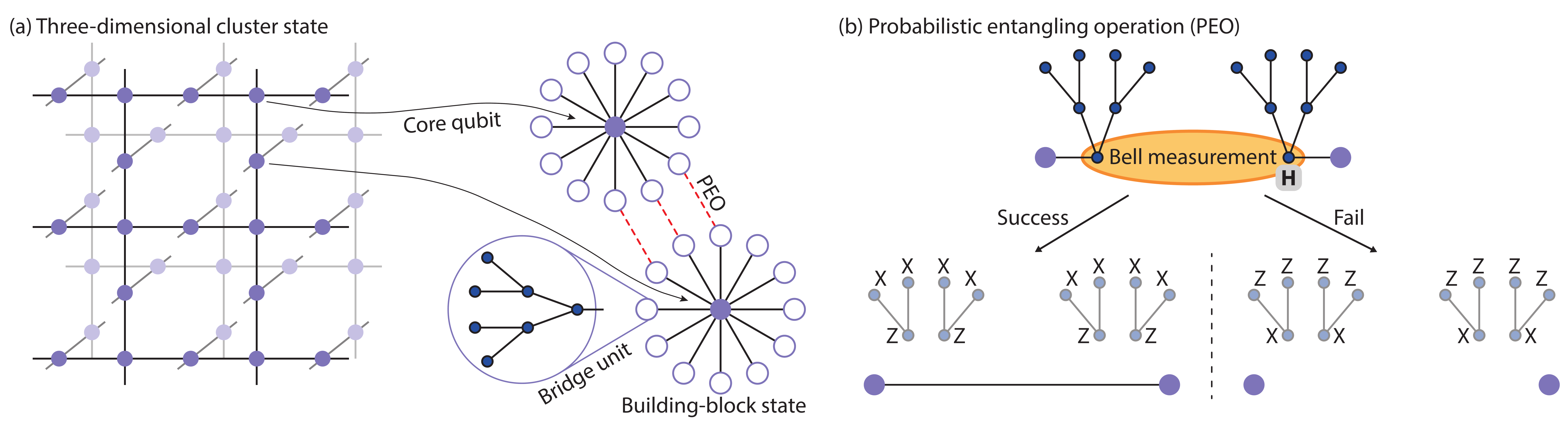}
\caption{
Protocol for linear optical quantum computing using 3D cluster states. This figure uses the {\it graph state} notation: each diagram represents the multi-qubit state which {\it would} result if one could prepare one qubit in state $\ket{+}$ for each node (filled circle), and then perform a controlled-phase gate for each edge (i.e. connecting line). However since LOQC does not permit deterministic entanglement, our states must be actually be created by a more lengthly process. (a) The full 3D cluster state is constructed in a near-deterministic step once we have created sufficiently complex building-block states. Each building-block state contains a photonic core qubit (solid circle) and several bridge units (empty circles), which are themselves tree graph states of photonic qubits. The core qubits of the building-block states will form the qubits in the 3D cluster state. Entanglement between these cores (i.e. edges in the eventual cluster state) is established by attempting probabilistic entangling operations (PEOs) between bridge units. (b) PEOs on two bridge units are used to make entanglement links between core qubits. The PEO is composed of a Hadamard gate on one of the bridge's root qubit followed by a Bell measurement (BM) between the two root qubits. Other photonic qubits in the bridge qubit are then measured in bases according to the outcome of the BM. If the BM succeeds, an entanglement link is generated. Regardless of whether it succeeds or fails, the remains of the two bridge units must be removed from their building blocks. 
}
\label{fig:protocol}
\end{figure*}

In this paper, we propose a protocol for LOQC that includes every step from the initial generation of entanglement primitives to the deployment of a fully fault-tolerant scalable unit for quantum computing. We consider computational errors and losses at each stage and endeavour to employ the most efficient known protocols for optical quantum information processing. In contrast to previous studies of noise thresholds in optical quantum computing~\cite{Dawson2006,HM2010}, we explicitly account for the substantial resource costs of LOQC protocols. We focus on a purely linear optical network, without employing matter qubits as memories or for entanglement generation. However we do assume the availability of on-demand sources of single photons, without concerning ourselves with the particular method with which these would be generated~\cite{Eisaman2011}. It is important to recognise that our results only represent an upper bound on the physical characteristics that are required of the components in an LOQC system -- our protocol can admit various further optimisations, and these will make the physical requirements less stringent. Nevertheless we believe the results we offer are highly relevant to the field, having been derived from protocols that are presently ``state of the art", and moreover it is therefore fair to compare the results here with those that have been derived for matter and hybrid matter-optical systems. 

Our analysis allows us to make an estimate of the overall scale of the resources necessary to construct a fully fault-tolerant optical quantum computer. We choose the number of detectors as a metric for the device size, recognising that the total numbers for the other kinds of component will scale roughly proportionately. We find that one would require upwards of $10^5 \text{ --- } 10^6$ detectors per physically encoded qubit in the cluster state, therefore requiring a total of at least $10^{11} \text{ --- } 10^{12}$ detectors to build a $1000$ logical qubit quantum computer~\footnote{This assumes that each logical qubit will be encoded in a surface code consisting of $\sim 1000$ physical qubits~\cite{Fowler2012}.}. Further, such a quantum computer would require loss rates per component below $\sim 10^{-3}$ and error rates below $\sim 10^{-5}$ per component. 

\section{Protocol}
\label{sec:protocol}

\begin{figure*}[!]
\centering
\includegraphics[width=0.75\linewidth]{\figpath 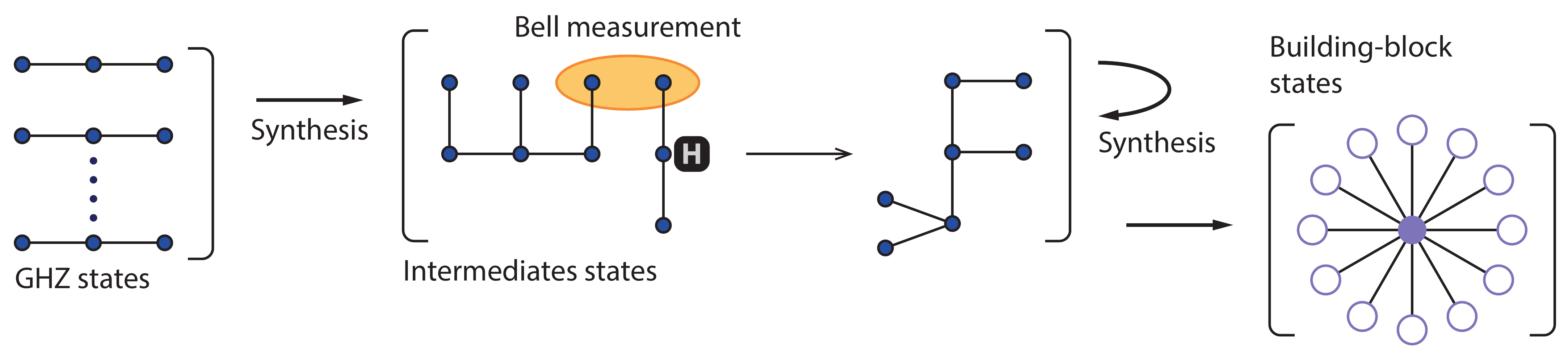}
\caption{
Building-block states are constructed in a series of stages from initial three-qubit GHZ state entanglement resources. Further details are given in Appendix~\ref{sec:AppConstruct}. 
}
\label{fig:buildingblock}
\end{figure*}

Our protocol is based on a three-dimensional (3D) cluster state~\cite{Raussendorf2006,Raussendorf2007PRL,Raussendorf2007NJP} [Fig.~\ref{fig:protocol}~(a)]. With the cluster state approach, all entanglement required by the quantum computation is generated ahead of the computation itself, which then proceededs purely through measurements. The 3D cluster states enables measurement-based implementations of topological quantum computing using the surface code~\cite{Kitaev2003,Dennis2002,Fowler2009}, providing high thresholds for both qubit loss and computational errors~\cite{Barrett2010}. Without qubit loss, 3D cluster states tolerate phase errors with a rate up to $3\%$ on each qubit; conversely, without computational errors, they tolerate up to $24.9\%$ qubit losses~\cite{Barrett2010}; and with both computational errors and qubit loss, the threshold of errors decreases approximately linearly with the loss rate. Thus cluster states are particularly well suited to LOQC as they can be efficiently prepared with linear optics: there is no fundamental difficultly caused by a high rate of entanglement failure during the creation of the cluster state, provided that once it is created it surpasses these thresholds~\cite{Nielsen2004,Browne2005,GS2014,Zaidi2014}. 

The 3D cluster state is a graph state on a 3D lattice, which can be understood by supposing that each vertex on the graph denotes a qubit initialised in the state $\ket{+}$ and each edge denotes an controlled-phase gate entangling the two linked qubits. In the particular lattice we require, each qubit is connected to four neighbouring qubits, see Fig.~\ref{fig:protocol}(a). In order to create such a cluster state, our protocol requires one complete building-block state to be prepared for each eventual qubit in the cluster. Importantly, these building-block states contain sufficient redundant encoding that entanglement links between building blocks can be generated with a probability above that necessary for fault-tolerant computing. If suitable building blocks can be constructed, a fault-tolerant cluster state of arbitrary size can then be generated deterministically. We can therefore focus on the optimal approach to constructing these building-block states (Fig.~\ref{fig:buildingblock}), without concerning ourselves with the precise details of the 3D cluster state that will ultimately be generated. However, we note that it is only necessary to generate one 2D layer of the cluster state ``at a time'', entangling it with the layer generated previously (each layer represents one `clock cycle' of the computation, and therefore requires one vertex for every physical qubit). Therefore the building-block factories can be reused to generate each layer of the 3D cluster. 

\begin{figure*}[!]
\centering
\includegraphics[width=0.95\linewidth]{\figpath 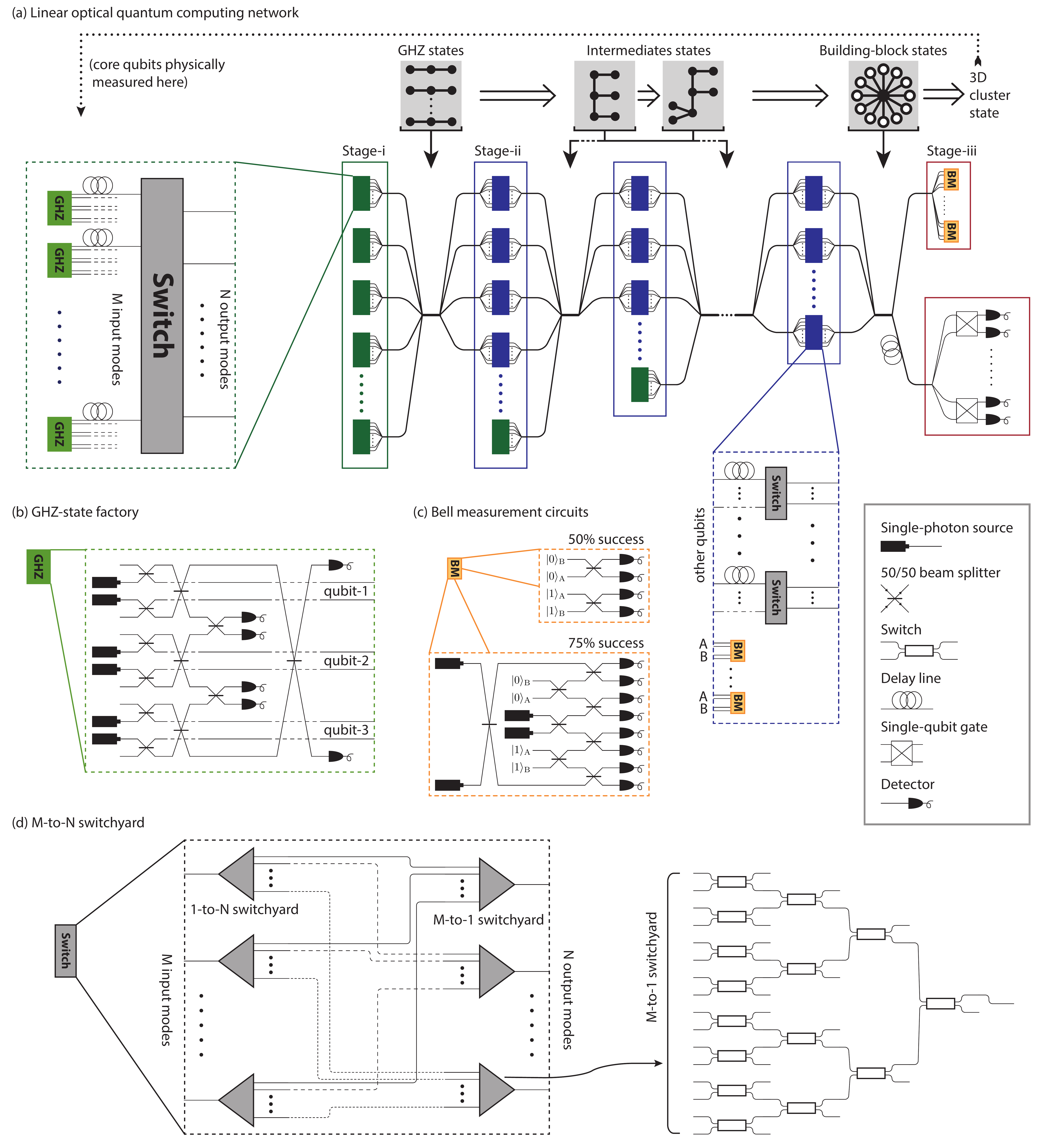}
\caption{
(a) Circuit for realising linear optical quantum computing using a three-dimensional cluster state. The circuit includes stages for (i) generating initial three-qubit GHZ states, (ii) synthesising these states into building-block states, and finally (iii) constructing the 3D cluster state. Qubits on the cluster state are physically measured right after they are generated by a GHZ state factory. (b) GHZ states factories probabilistically generate three-qubit GHZ states from six single photons. Successful generation of the GHZ state is heralded by specific three photon detection events at the detectors. To select at most $N$ successful copies of the GHZ state from $M$ attempts, we need six $M$-to-$N$ switchyards. Delay lines are necessary before photons enter switchyards to allow for feed-forward. (c) Bell measurement circuits synthesise building-block states. The two different circuits depicted succeed with probabilities $50\%$ and $75\%$, respectively. For the $75\%$-success circuit, four ancillary single photons are used. At each synthesise stage, many copies of input states are prepared, and measurements are performed on these states in parallel. Successful output states are selected with switch networks. (d) A $M$-to-$N$ switchyard can be realised with $1$-to-$N$ switchyards and $M$-to-$1$ switchyards. For each $1$-to-$N$ switchyard, every output mode is connected to an input mode of a different $M$-to-$1$ switchyard. A $M$-to-$1$ switchyard composed of $\sim 2M$ $2$-to-$2$ switches. A $1$-to-$N$ switchyard is similar. 
}
\label{fig:circuits}
\end{figure*}

As in other cluster state generation protocols~\cite{Nielsen2004,Barrett2005,Benjamin2005,Duan2005,Bodiya2006,Matsuzaki2010,Li2010}, our building-block state is also a graph state. We employ the star graph as the basic structure of our building blocks, Fig.~\ref{fig:protocol}~(a). This state is composed of one core qubit and several bridge units. While the core qubit is a single photonic qubit, each bridge unit is physically encoded in tree-structure graph state of several photonic qubits. In order to implement a PEO between two different building block states, a Bell measurement is carried out between the root qubits at the base of each bridge unit. The tree-like structure within the bridge units enables two key properties: In the case of a successful PEO between bridge units on two different building-block states, the core qubits on each building block become entangled, and the remaining qubits within each unit can be trimmed away, see the left part of Fig.~\ref{fig:protocol}~(b). Moreover, on failure of the entangling operation then the measurements on the remaining qubits within the bridge units allow us to identify any necessary phase correction to the core qubit, preventing its corruption (with high probability). The right part of Fig.~\ref{fig:protocol}~(b) summarises this protocol, see Appendix~\ref{sec:AppLossAndErrorTolerance} for further details. This method for recovering from PEO failure via measurements on ancillary qubits follows the approach introduced in Ref.~\cite{Varnava2006}.

With these two properties, it is possible to make multiple attempts to form links between core qubits while still ensuring that errors remain below the fault-tolerant threshold. Two building blocks can therefore be successfully connected with a high rate provided that there are enough bridge units. 

Since each core qubit must be linked to four other core qubits, the number of bridge units on the building blocks is chosen to be a multiple of four, with a quarter of the bridge units allocated for each connection. To establish an entanglement link, PEOs are performed on corresponding bridge unit in parallel. If there is one successful PEO and the removal measurements are also successful, then the connection is successfully established. If there is more than one successful PEO, we keep only one link. The connection fails if there are no successful PEOs or one of the removal operations fails. Connection failures are dealt with by treating core qubits with a failed connection as missing qubits~\cite{Li2010}, which can be tolerated by MBQC on the 3D cluster state. 

The finally prepared state of core qubits is equivalent to the 3D cluster state up to some feedback single-qubit gates depending on outcomes of single-qubit measurements and Bell measurements for preparing the state. In the MBQC algorithm on the 3D cluster state, (core) qubits are measured in four bases, which are $\sigma^x$, $\sigma^z$, and $(\sigma^x \pm \sigma^y)/\sqrt{2}$ (only for magic state injection). In our protocol, the feedback gate on a core qubit is always either the identity $\openone$ or the phase gate $\sigma^z$. Therefore, remarkably, core qubits can be measured before the cluster state is prepared! It is beneficial to do so, in order to reduce the effect of photon loss: core qubits are measured as early as possible, and obviously if a measurement fails then that particular building block is abandoned at its initial stage. Once the full feedback is known, we may update (flip) the recorded outcomes of any cores measured in $\sigma^x$ or $(\sigma^x \pm \sigma^y)/\sqrt{2}$. 

\section{Generation of building blocks}

\begin{figure*}[tbp]
\centering
\includegraphics[width=1\linewidth]{\figpath 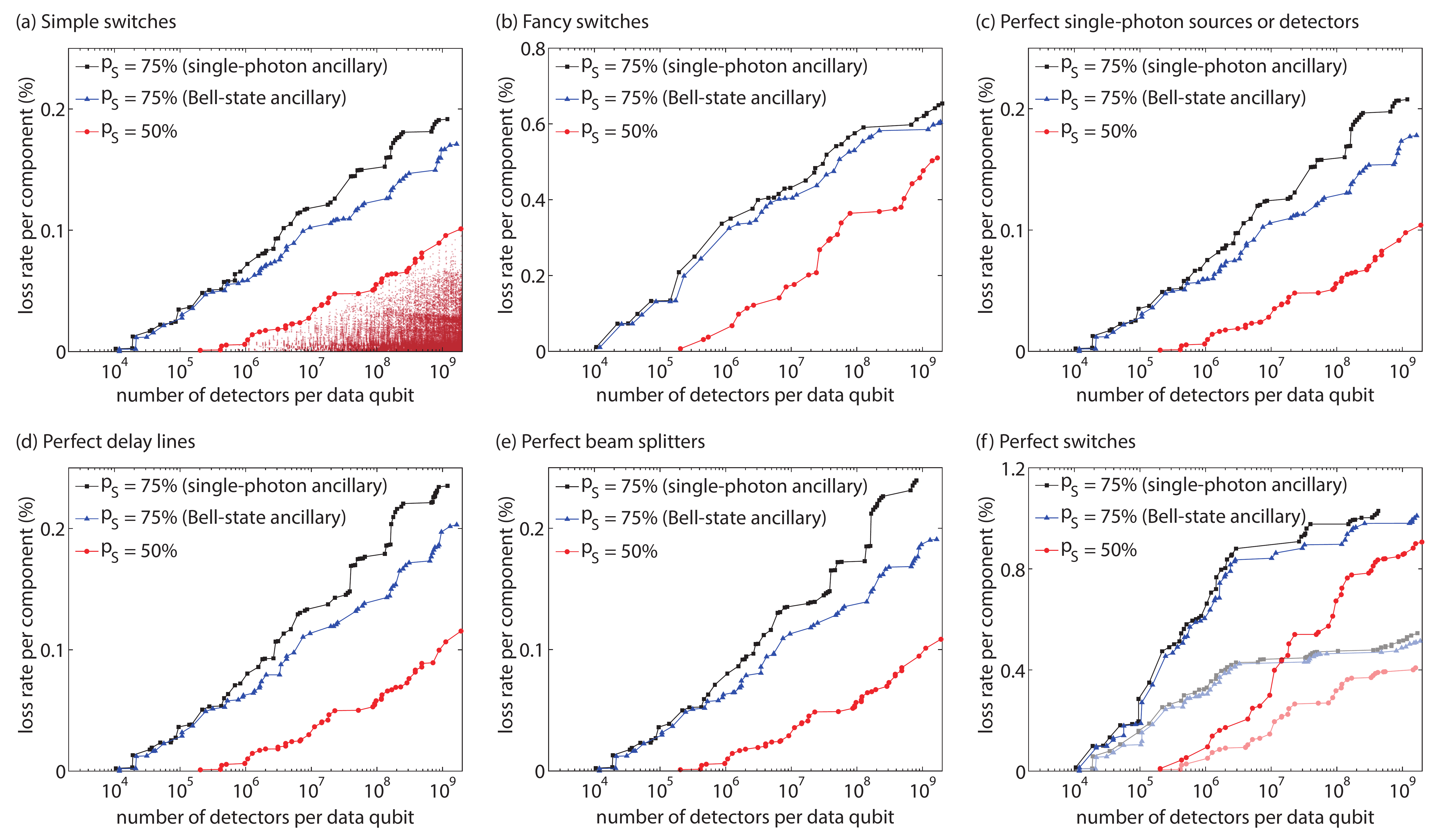}
\caption{
Thresholds on the loss rate per component $p$ as a function of the number of detectors per data qubit in the case of no computational errors; $\pro{S}$ denotes the success rate of Bell measurements. We again note that each logical qubit will be encoded in a surface code consisting of $\geq 1000$ data qubits. (a) Model in which each switchyard is composed of cascaded $2$-to-$2$ switches. All components have the same loss rate, i.e.~$\pro{e} = \pro{b} = \pro{d} = \pro{s} = \pro{m} = p$. (b) Each switchyard is a single switch with multiple inputs and outputs. All components have the same loss rate $p$. (c),~(d),~(e)~and~(f) In each plot, one form of component is assumed to be perfect, while all other components have the same loss rate $p$. As can be seen, the most dramatic improvement is seen when the switches are assumed to be perfect. Transparent curves in subfigure (f) corresponds to the case that the loss rate of switches is $10\%$ of other components. 
}
\label{fig:lossrate}
\end{figure*}

Each building-block state must be generated from an initial resource of unentangled single photons. In our scheme, these single photons are first entangled into three-qubit GHZ states. These entanglement primitives can then be sequentially combined into larger units using Bell measurements (Fig.~\ref{fig:buildingblock}). This process is known to be efficient for loss rates of less than 1/3~\cite{Varnava2008}, since at each stage it is then possible increase (up to double) the size of the resulting entangled states. Further details on the building-block construction process are given in Appendix~\ref{sec:AppConstruct}. 

Regardless of the specific architecture of the building block to be generated, this process requires two primary circuit elements. The first element, a GHZ-state factory, produces GHZ states probabilistically from single-photon inputs. The second element probabilistically joins two independent graph states into a larger graph state. Along with these two processing elements, it is also necessary to construct switching networks and delay lines in order to route photons between processing states. All of these operations must be realised using only linear optical elements, e.g.~single-photon sources, beam splitters, switches, delay lines and photon detectors [Fig.~\ref{fig:circuits}~(a)]. 

In our scheme, we use the same GHZ-state factory as proposed in Ref.~\cite{Varnava2008}. This circuit requires six single photon inputs, and, in the lossless case, successfully generates GHZ states with probability $1/32$ [Fig.~\ref{fig:circuits} (b)]. Our fusion elements use Bell measurements as PEOs for joining intermediate states. These Bell measurements consume one photon from each input state~\cite{Browne2005}. A tempting alternative is to employ the Type-I fusion gate, which consumes only one photon and can also connect two graph states~\cite{Browne2005}. However, a Type-I fusion gate may convert photon loss into computational errors (see Appendix~\ref{sec:AppConstruct}), which should be avoided as overcoming errors is usually harder than overcoming photon loss. Therefore, we only use Bell measurements in our protocol~\cite{Varnava2008}. The circuits we use for the Bell measurement are also shown in Fig.~\ref{fig:circuits}~(c). Without any ancillary resources, a linear optical Bell measurement (often termed Type-II fusion) can succeed with $50\%$ probability in the lossless case. However, with the help of four ancillary single photons, the success probability of a Bell measurement can be boosted to $75\%$~\cite{Grice2011}. The same success probability can also be achieved with a Bell-state as the ancillary resource~\cite{Ewert2014}. With a resource state of more entangled photons, the success probability can be further boosted~\cite{Grice2011,Ewert2014}. 

As neither GHZ-state generation nor Bell measurements can succeed deterministically, we select successful outcomes from these operations to feed into the next stage of construction. This requires a rapidly reconfigurable switchyard consisting of a network of switches. For example, to select $N$ successful copies of the three-qubit GHZ state from $M$ attempts in parallel, we need six $M$-input to $N$-output switchyards, one for each output mode of the GHZ-state generation circuit. Before photons enter switchyards, delay lines are necessary to allow time for the switchyard to be reconfigured. 

We consider two different approaches to this switching requirement. In the ideal case, this switchyard would consist of a single reconfigurable switch with multiple inputs and outputs~\cite{Kim2003}, in which there is no extra cost in terms of losses or errors as $N$ or $M$ increases. This may prove impossible to achieve, and so we also consider the opposite limit, in which a switchyard is built out of a network of $2$-to-$2$ switches. Such an $M$-to-$N$ switchyard can be realised with $M$ $1$-to-$N$ switchyards and $N$ $M$-to-$1$ switchyards [Fig.~\ref{fig:circuits}~(d)]. Each $1$-to-$N$ and $M$-to-$1$ switchyard is respectively composed of $\sim N$ and $\sim M$ $2$-to-$2$ switches, as also shown in Fig.~\ref{fig:circuits}~(d). With such a network of simple switches, each photon must go through approximately $\log_2 (MN)$ switches. To minimise photon loss, switchyards with multiple inputs $M$ but a single output $N$ are favourable. However resources are not used efficiently in this case, and many successful PEO outputs will be discarded. To increase the efficiency, it is preferable to use switchyards with more output modes. In our numerical simulations, we have considered different configurations of the switch network to obtain the optimal threshold of a computer built with $2$-to-$2$ switches. 

\section{Photon loss and computational errors}

The main source of noise in LOQC is photon loss, which may be induced by any component on the optical path of the qubit. We assume that a loss occurs at single-photon sources, beam splitters, delay lines (for the time period required for one PEO stage), switches and detectors with the rates $\pro{e}$, $\pro{b}$, $\pro{d}$, $\pro{s}$ and $\pro{m}$, respectively. 

In addition to photon losses, we also have to consider computational errors. Because measurements are eventually attempted on all photonic qubits in the protocol, computational errors are induced by any source of noise that can affect these measurement outcomes. For example, any asymmetry, e.g.~phase difference or biased transmission, between two modes of a qubit may result in computational errors. All computational errors are equivalent to Pauli errors (see Appendix~\ref{sec:AppCEM}). In this paper, we assume that depolarising errors may happen at beam splitters, delay lines and switches with the rates $\eps{b}$, $\eps{d}$ and $\eps{s}$, respectively. Imperfect modal overlap between different photon sources will lead to imperfect quantum interference at beam splitters, and therefore also to Pauli errors. For simplicity, in our model we incorporate this form of error into $\eps{b}$. 

Other types of noise are also tolerable in our protocol. For example, a photon source may emit two photons rather than a single photon into the circuit. Similar errors can be induced by switching errors, in which a photon enters the wrong mode, and from dark counts of detectors. To first order, all of these errors will be caught during measurement, since if these extra photons survive in the optical path we will measure more than the expected number of photons. In this case, we can simply treat the qubit as missing, an error which can be overcome in the same way as true photon loss. However, if a two-photon error is followed by a photon loss event, only one photon will be detected, and the measurement on such a qubit may give a wrong outcome. These computational errors are also equivalent to Pauli errors and can be corrected with our protocol. Although these errors can be corrected, we consider regimes in which they will occur at a rate much lower than the first-order error terms, and so we do not explicitly include them in our threshold study. 

\section{Thresholds}

In this approach to LOQC, the fault-tolerance threshold depends on the complexity of each building-block state. With more resources, one can prepare bigger building blocks, and thus a higher level of photon loss is tolerable. In Fig.~\ref{fig:lossrate}, fault-tolerant thresholds are obtained numerically (see Appendix~\ref{sec:AppNS} for details). In order to provide some physical intuition for the size of such linear optical quantum computers, we choose to specify the total number of detectors needed as our metric of the resources required. It can be seen that this approximately corresponds to twice the number of single photons needed, and therefore twice the number of single photon sources. It is likely that the resource burden of the other elements, e.g.~beam splitters, delay lines and switches, will be of similar magnitudes. 

Note that each curve in Fig.~\ref{fig:lossrate}~(a) is actually an envelope representing the best of a very large number of protocols that were tested. Each small dot within the red curve in the upper left figure represents the outcome of one such simulation; these dots are omitted from other curves for clarity. 

It is vital to appreciate that the number of detectors shown in the figure is for a single building-block state rather than the entire computer. A building-block state corresponds to only one qubit on the cluster state, i.e.~one data qubit of the surface code, which could correspond to just one ion in an ion trap quantum computer or one superconducting qubit in a superconducting quantum computer. It is anticipated that a fault-tolerant quantum computer will need at least $\sim 10^6$ data qubits in order to be able to compete with state-of-the-art classical computers~\cite{Fowler2012,Hamdi2014}. We therefore do not consider building-block states with a resource requirement of greater than two billion detectors, since at that point one finds the entire computer requires thousands of trillions of components! 

In Fig.~\ref{fig:lossrate}~(a), we consider the case in which all components of the computer have the same photon loss rate. Depending on the choice of Bell measurement protocol, the threshold loss rate per component varies from $\sim 0.1\%$ to $\sim 0.2\%$. In this subfigure, we consider the worst case approach, in which each switchyard is built out of a cascade of $2$-to-$2$ switches. For comparison, in Fig.~\ref{fig:lossrate}~(b) we consider a more sophisticated computer, in which each switchyard is a multiple-input multiple-output switch with the same loss rate as the other components. In this case the threshold is $\sim 3 \text{ --- } 5$ times higher than that of a simple-switch computer. We note that these thresholds approach the $1\%$ limiting loss rate that has been discussed in the context of a computing paradigm where gates are essentially {\it deterministic} but suffer a small probability of qubit loss~\cite{Whiteside2014} (we have achieved this at the cost of the additional resource overhead of course).

In order to further explore which components have the most significant impact on the fault-tolerance threshold, in Figs.~\ref{fig:lossrate}~(c),~(d),~(e)~and~(f) we modify the model in Fig.~\ref{fig:lossrate}~(a), assuming in each that one of the circuit components is lossless. These simulations confirm that it is the switching networks which most strongly impact the loss tolerance of the quantum computer. This suggests that alternative approaches in which intermediate cluster states are extensively recycled (similar to the recycling discussed in Ref~\cite{Matsuzaki2010}) will suffer from the associated increase in complexity of the switching networks. 

The threshold changes dramatically with increased success probability Bell measurements. With a higher success probability, the size of building-block states (number of bridge units) can be smaller, hence both the level of noise and the resource cost can be lower. Bell measurements with $75\%$ success probabilities perform significantly better than those with $50\%$ success probability. Further, the single-photon ancilla assisted Bell measurement is slightly better than the Bell-state assisted Bell measurement. These boosted Bell measurements do however require photon detectors with additional photon-number resolution. In the case of the $50\%$-success Bell measurement, we need detectors that can distinguish photon numbers $0,1,2$, while for $75\%$-success Bell measurements, we need detectors that can distinguish photon numbers $0,1,2,3$. With more complex ancillary states, the success probability can be further boosted. However in this case more resources are required for preparing these ancillary states, which can counteract the benefits of the higher success probabilities (see Appendix~\ref{sec:AppBM}). 

\begin{figure}[tbp]
\centering
\includegraphics[width=0.93\linewidth]{\figpath 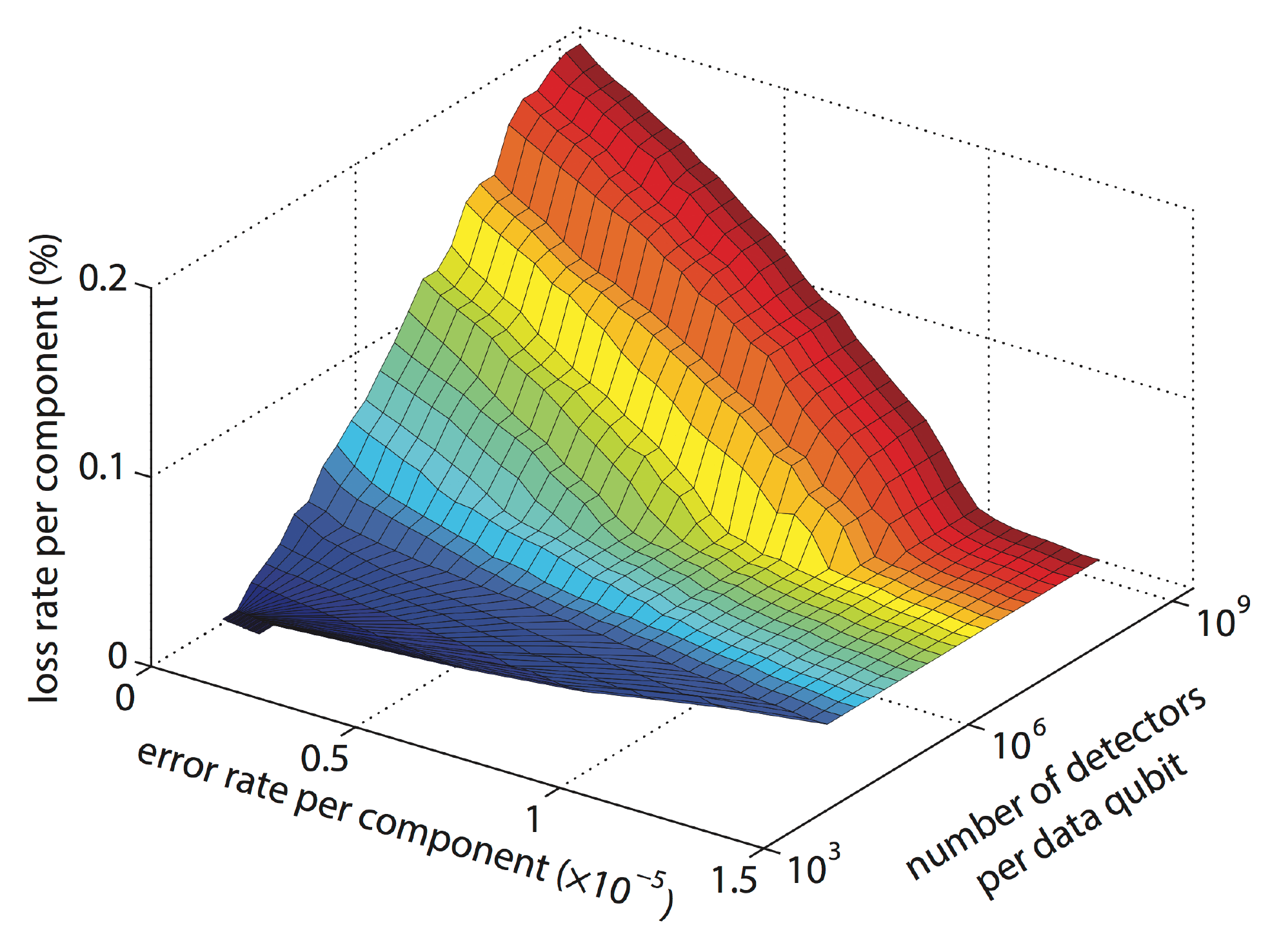}
\caption{
Thresholds on the loss rate per component $p$ for given the error rate per component $\epsilon$ and the number of detectors per data qubit. Switchyards are composed of cascaded $2$-to-$2$ switches and single-photon ancilla assisted Bell measurements with $75\%$ success probability are used. All components have the same loss rate, i.e.~$\pro{e} = \pro{b} = \pro{d} = \pro{s} = \pro{m} = p$ and all error rates are equal, i.e.~$\eps{b} = \eps{d} = \eps{s} = \epsilon$. See Fig.~\ref{fig:error} for the full data with more details. 
}
\label{fig:3D}
\end{figure}

The presence of computational errors reduces the threshold loss rate. A general study of the threshold for both loss rates and error rates is shown in Fig.~\ref{fig:3D}, in which we only consider quantum computers built with $2$-to-$2$ switches and single-photon ancilla assisted Bell measurements with $75\%$ success probability. More data for other Bell-measurement circuits and fancy switches can be found in Fig.~\ref{fig:lossrate-error}. The threshold error rate per component is on the order of $10^{-5}$. 

\section{Discussion}

We have proposed a comprehensive protocol for LOQC with 3D cluster states, in which we consider the full network of linear optical devices necessary to realise a quantum computer. We find thresholds for loss and error rates of $\sim 10^{-3}$ and $\sim 10^{-5}$ per component, respectively. This per-component performance is  beyond the current state of the art in photonics~\cite{Ringbauer2014d,Harris2014,Goh2001,Shadbolt2011a,Niclass2008, Miki2014,Yao2011}. Furthermore we find that such a quantum computer would require on the order of $10^{11}$ detectors, and similar numbers of other components including deterministic and indistinguishable single photon sources. These component counts are several orders of magnitude greater than those required for systems with deterministic gates~\cite{Fowler2012,BenjPRX}. 

We wish to emphasise that these stringent thresholds should be taken as a challenge to the community, aiming to stimulate further discussion and innovation in LOQC. From an experimental perspective, we have tried to determine which components will prove most critical in the development of an optical quantum computer. We found that, for our scheme, it is the performance of the optical switches that have by far the most impact the threshold loss and error rates, while other components contribute more equally. We hope that this will help guide the priorities of future experimental work aimed towards realising LOQC. From a theoretical perspective, we hope that our work will stimulate others to improve on our thresholds by exploring alternative schemes. 

\begin{acknowledgments}
This work was supported by the EPSRC platform grant `Molecular Quantum Devices' (EP/J015067/1), and the EPSRC National Quantum Technology Hub in Networked Quantum Information Processing. We would like to thank Joshua Nunn, Animesh Datta, Benjamin Metcalf, Jacques Carolan, Dan Browne and Mercedes Gimeno-Segovia for helpful discussions. 
\end{acknowledgments}

\newpage
\newpage

\appendix

\section{Loss and error tolerant building blocks}
\label{sec:AppLossAndErrorTolerance}

\begin{figure*}[tbp]
\centering
\includegraphics[width=0.74\linewidth]{\figpath 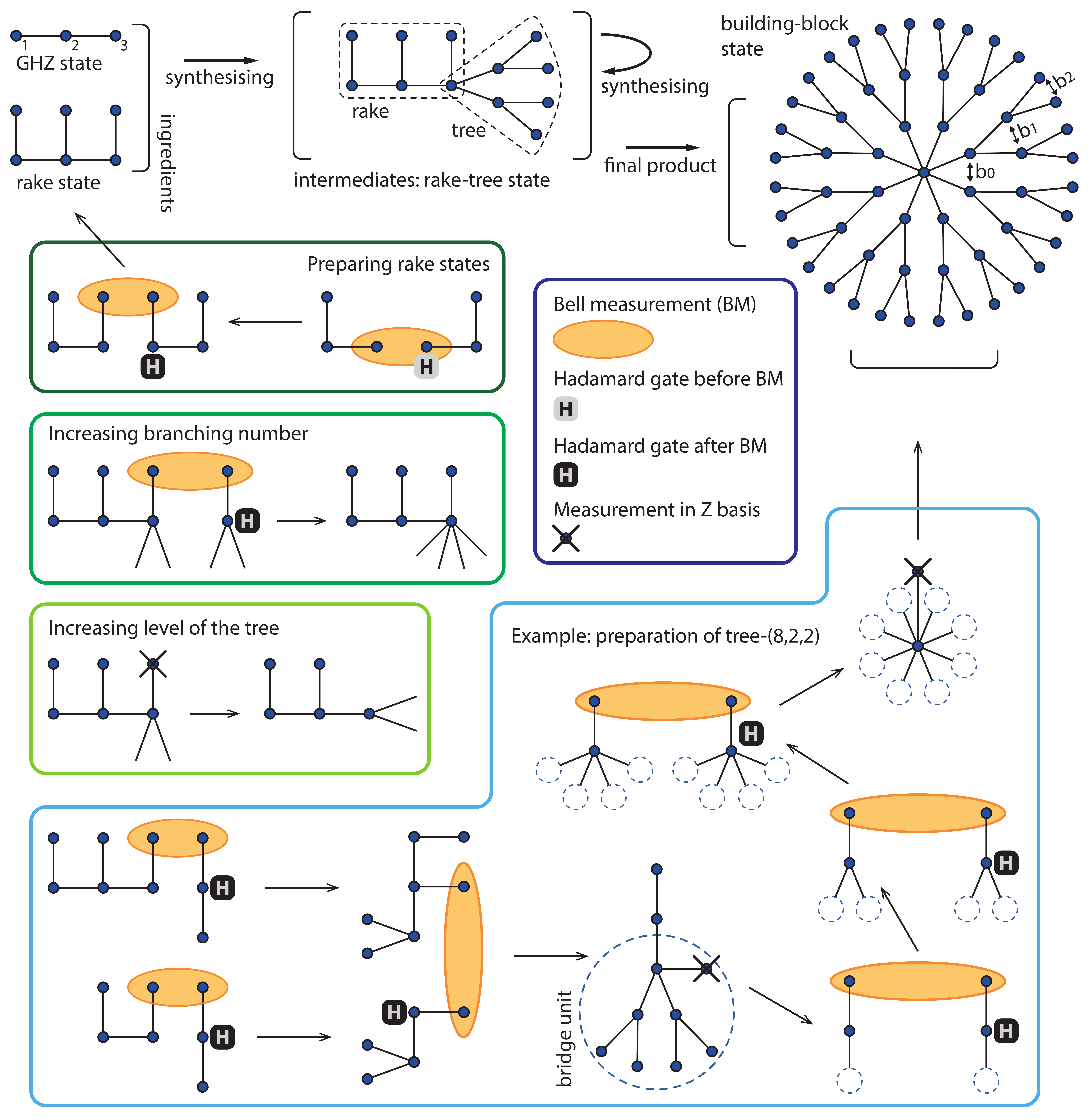}
\caption{
Scheme for generating building-block states with three-qubit GHZ states. The building-block state is a tree graph state of photonic qubits. Such a tree can be characterized by branching numbers $(b_0,b_1,b_2,\ldots)$, where $b_0$ is the number of bridge units, and $b_i$ is the number of $i$th-generation branches of a bridge unit. In our protocol, $b_0$ is always a multiple of four. Graph states occurring in the generation process include three-qubit GHZ states, rake states, tree states, and rake-tree states. On the top of the figure, the example rake state has $3$ branches, the rake-tree state is composed of a rake with $3$ branches and a $2$-level tree with branching numbers $(2,2)$, and the building-block state is a $3$-level tree state with branching numbers $(8,2,2)$. 
}
\label{fig:preparation}
\end{figure*}

As discussed in the main text, our protocol is based on a three-dimensional (3D) cluster state [Fig.~\ref{fig:protocol}~(a)]. To create the cluster state, one intermediate building-block state must be prepared for each qubit in the cluster. 

We employ the star graph as the basic structure of our building blocks [Fig.~\ref{fig:protocol}~(a)]. This state is composed of one core photonic qubit and several bridge units. To tolerate photon loss and failures of PEOs, each bridge unit is encoded as a tree-structure graph state of several photonic qubits with a root qubit connected to the core qubit [Fig.~\ref{fig:preparation}]. The PEO for connecting two core qubits includes a Hadamard gate on one root qubit and a Bell measurement on two root qubits [Fig.~\ref{fig:protocol}~(b)]. Because the Bell measurement can only succeed probabilistically in LOQC, the overall operation is probabilistic. If the PEO is successful (fails), qubits on first-generation branches, which are directly connected to the root, are measured in the $\sigma^z$ ($\sigma^x$) basis, qubits on second-generation branches are measured in the $\sigma^x$ ($\sigma^z$) basis, and so on. This measurement pattern removes redundant branches from two connected building blocks if the PEO is successful and removes entire bridge units from two independent building blocks if the PEO is failed. This removal operation is not always successful due to photon loss. When the tree graph state is large enough, the removal operation can succeed with an arbitrarily high probability if the photon loss rate is lower than $50\%$~\cite{Varnava2006}. 

\section{Constructing building blocks}
\label{sec:AppConstruct}

\begin{figure}[tbp]
\centering
\includegraphics[width=1\linewidth]{\figpath 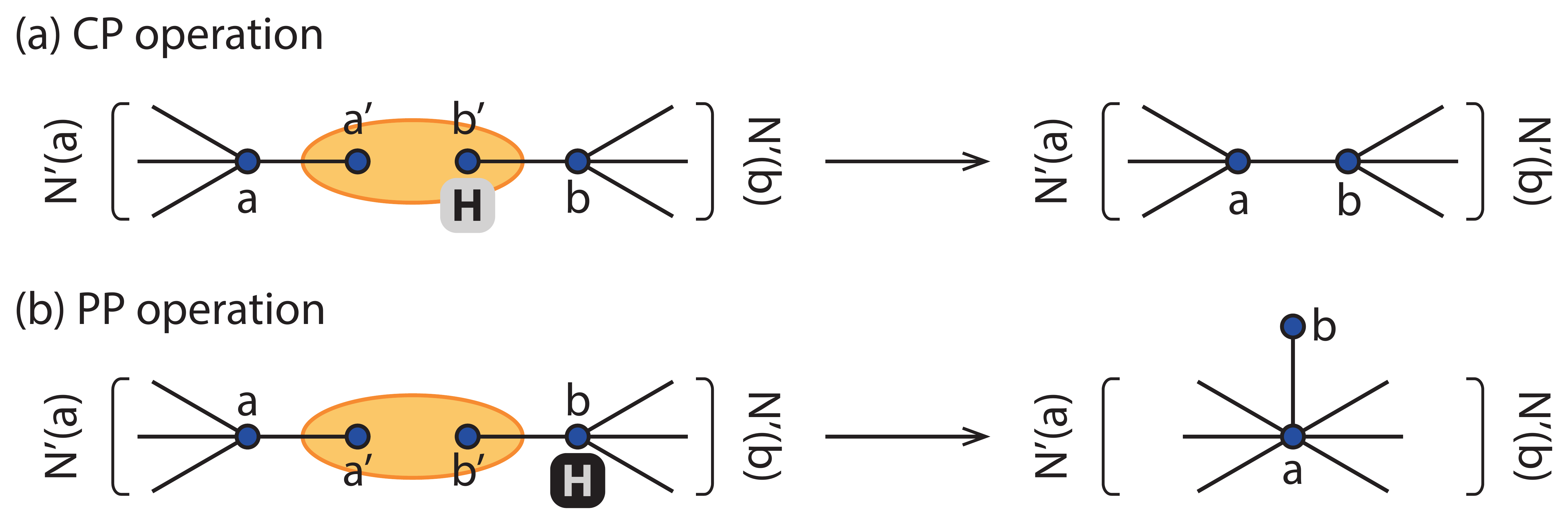}
\caption{
Fusion operations on graph states using Bell measurements in the basis $\ket{\text{BS}_{\mu,\nu}} = Z_{a'}^\mu X_{b'}^\nu (\ket{00}_{a',b'} + \ket{11}_{a',b'})/\sqrt{2}$, where $\mu,\nu=0,1$ (a) A controlled-phase (CP) operation includes a Hadamard gate on one of two measured qubits and then the Bell measurement. Depending on the measurement outcome, an operation $Z_a^\nu Z_b^\mu$ need to be performed. (b) A parity-projection (PP) operation includes the Bell measurement and a Hadamard gate on the nearest neighbouring qubit of one of two measured qubits. Depending on the measurement outcome, an operation $Z_a^\nu Z_b^\mu \prod_{i\in N'(b)}Z_i^{\mu}$ need to be performed. Here, $N'(c)$ denotes neighbouring qubits of the qubit-$c$ except the qubit-$c'$. 
}
\label{fig:operation}
\end{figure}

Building-block states are generated by fusing three-qubit GHZ states (see Fig.~\ref{fig:preparation}) with Bell measurements. There are four types of graph states occurring in the generation process, which are three-qubit GHZ states, rake-structure graph states, tree-structure graph states, and rake-tree states. In the first step, rake states are prepared from GHZ states. These states, along with further GHZ states are the basic ingredients of rake-tree states. Using these ingredients, rake-tree states are generated and enlarged with Bell measurements. When the tree of a rake-tree state is large enough, it can be converted into a building-block state by removing the rake. As an example, the construction process for a building-block state with branching numbers $(8,2,2)$, is shown in Fig.~\ref{fig:preparation}. 

A rake with $r$ branches can be prepared with $2(r-1)$ GHZ states (assuming all Bell measurements are successful) in $\ceil[\log_2 (r-1)] + 1$ steps. In the first step, each pair of GHZ states are fused by a CP operation [see Fig.~\ref{fig:operation}~(a)] to obtain a $4$-qubit linear cluster state, which is also a rake with $2$ branches. Two rakes can be combined into a bigger rake by a PP operation [see Fig.~\ref{fig:operation}~(b)]: If two input rakes respectively have $r_1$ and $r_2$ branches, the output rake has $r_1+r_2-2$ branches. Since, in each step the number of branches can be nearly doubled, $r-1$ $2$-branch rakes can be combined into the $r$-branch rake in $\ceil[\log_2 (r-1)]$ steps. 

\begin{figure*}[tbp]
\centering
\includegraphics[width=1\linewidth]{\figpath 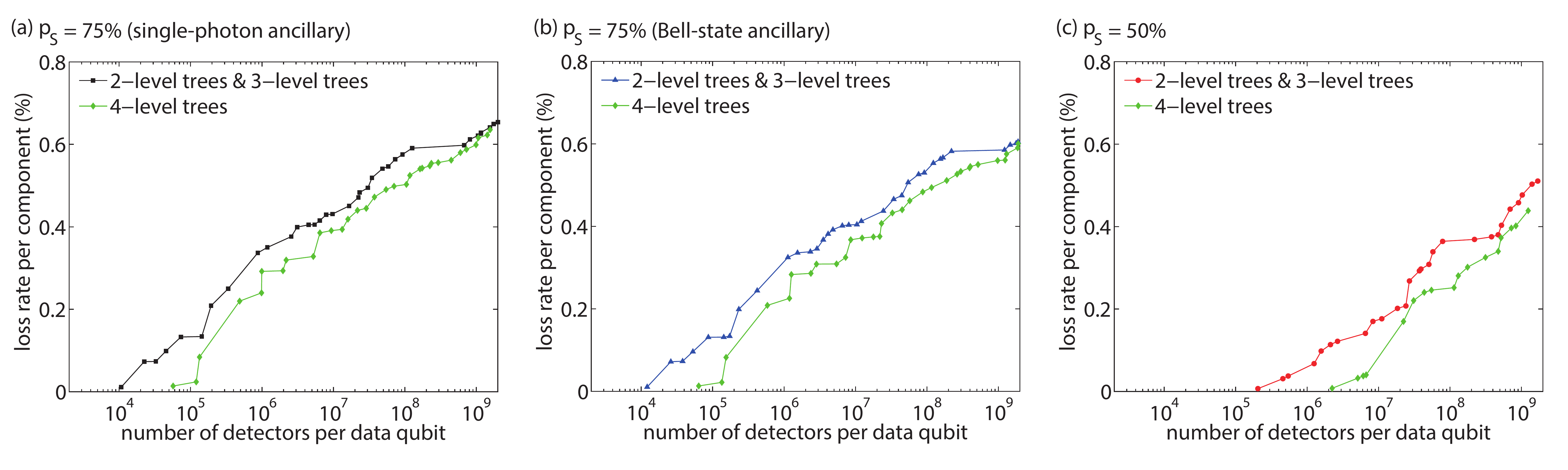}
\caption{
Thresholds of the loss rate per component without computational errors for a computer built with fancy switches. All components have the same loss rate. In addition to $2$-level trees and $3$-level trees, we have also considered $4$-level trees. One can find that $4$-level trees tolerate less photon loss per component within the regime we are interested. 
}
\label{fig:lossrate-trees}
\end{figure*}

A rake itself is a rake-tree graph state in which the tree is $1$-level but the branching number is $0$. A GHZ state itself is also a rake-tree graph state in which the rake has only $1$ branch and the tree is $1$-level with the branching number $1$. From these two kinds of graph states, rake-tree states can be generated and enlarged with two basic processes: increasing the branching number of the tree and increasing the level of the tree (see Fig.~\ref{fig:preparation}). The branching number is increased by fusing two rake-tree graph states with a PP operation, where one of the rake always has only one rake branch. The level of the tree is increased by removing, i.e.~measuring in the $\sigma^z$ basis, one branch of the rake. We would like to remark that, when the level number is increased from $1$ to $2$, the branch of the rake (which is supposed to be measured) can be kept as a branch of the tree. 

In our protocol of generating photonic tree-structure graph states, the rake structure allows us to increase the level of the tree with a single-qubit measurement (which is physically performed at the GHZ-state-generation stage due to the same reason of measuring core qubits as early as possible). This process is efficient; the largest rake state can be prepared in $\ceil[\log_2 (R-1)]+1$ steps, where $R$ is the level of the final tree. As a comparison, the previous protocol reported in Ref.~\cite{Varnava2008} requires probabilistic Bell measurements for increasing the level of the tree. Therefore, the number of construction stages is reduced in our protocol for high-level trees. Minimising the number of construction stages can reduce noise induced by delay lines and switchyards and also to reduce the resource cost. If the success probability of a Bell measurement is $\pro{S}$, for each successful output state, roughly speaking, $1/\pro{S}$ input states need to be prepared. For a GHZ state going through $n$ Bell-measurement stages, $1/\pro{S}^n$ copies will be required to ensure that each stage is successful. Reducing $n$ is therefore critical to reducing the the resource costs.

As we have discussed in the main text, we choose Bell measurement rather than Type-I fusion gate as the operation of entangling photons, because a Type-I fusion gate may convert photon loss into computational errors. For a Type-I fusion gate, if there is no photon loss, two input photonic qubits are projected into the subspace of HH and VV (for the polarisation encoding) when only one photon is detected, or the state HV (VH) if zero photons (two photons) are detected. With photon loss, the input qubits may be in the state VH rather than the subspace of HH and VV if only one photon is detected and the other is missing. Therefore, photon loss may result in computational errors in a Type-I fusion gate. 

\section{The model of computational errors}
\label{sec:AppCEM}

\begin{figure*}[tbp]
\centering
\includegraphics[width=0.67\linewidth]{\figpath 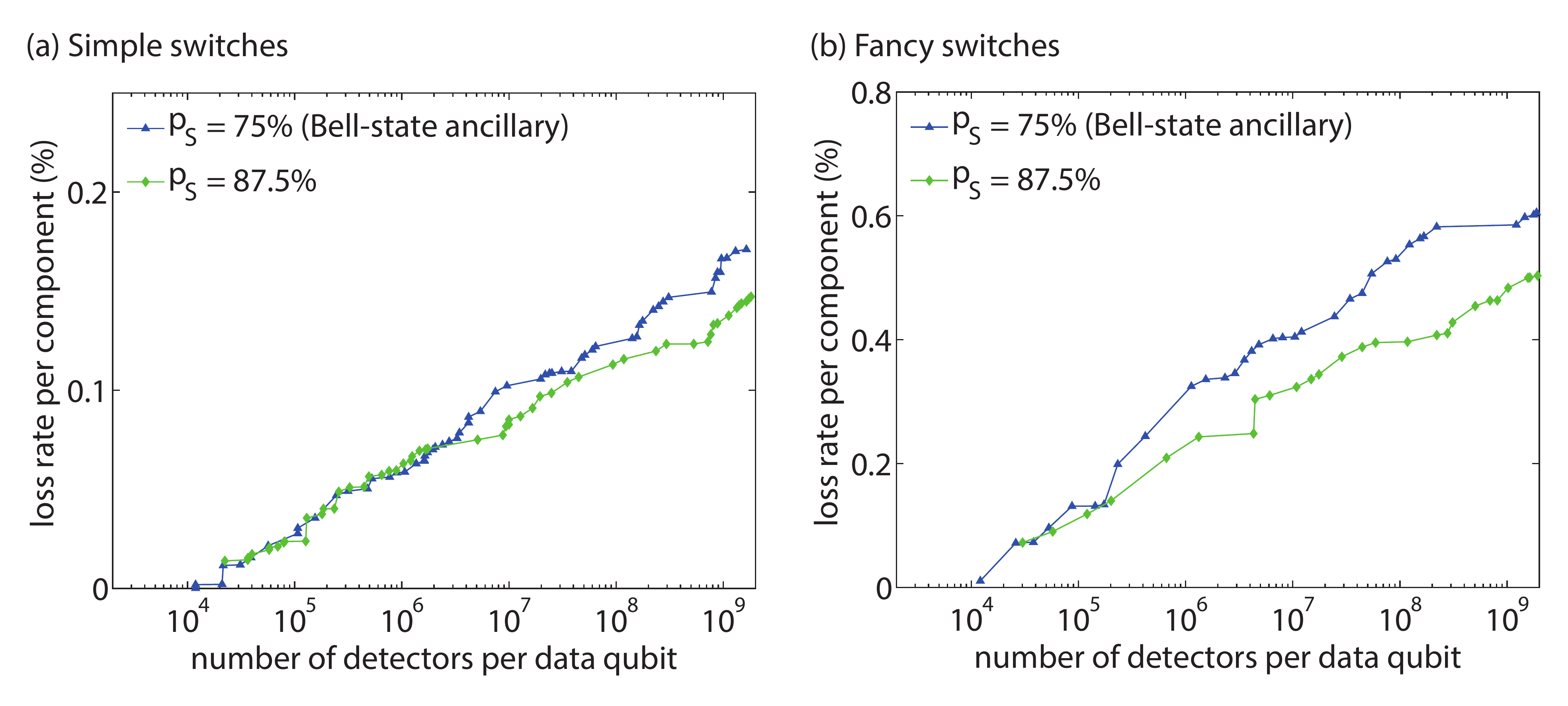}
\caption{
Thresholds of the loss rate per component without computational errors using Bell measurements with the success probability $75\%$ (assisted by a Bell state) and the success probability $87.5\%$ (assisted by a $4$-qubit GHZ state). All components have the same loss rate. 
}
\label{fig:lossrate-875}
\end{figure*}

In our protocol of LOQC, operations performed on three-qubit GHZ states include Hadamard gates, Bell measurements and single-qubit measurements in the $\sigma^x$ and $\sigma^z$ bases. Single-qubit measurements in bases $(\sigma^x \pm \sigma^y)/\sqrt{2}$ (for magic state injection) do not affect the fault-tolerance threshold. Therefore, any computational error is equivalent to a bit flip error $[\sigma^x]$, a phase flip error $[\sigma^z]$ or a combined error $[\sigma^y]$ on a single qubit, or a combination of these three types of errors on a group of qubits. Because GHZ states are prepared separately, there is no correlation between them right after they are generated. Correlations between GHZ states may occur when two qubits of different GHZ states are measured by a Bell measurement. However, for the Bell measurement on qubits $A$ and $B$, all errors are equivalent to three types of Pauli errors $[\sigma^x_A]$, $[\sigma^y_A]$ and $[\sigma^z_A]$ (or equivalently $[\sigma^x_B]$, $[\sigma^y_B]$ and $[\sigma^z_B]$, corresponding to three possible incorrect outcomes), which are all single-qubit errors. Therefore, all computational errors are equivalent to Pauli errors within GHZ states. For the GHZ state shown in Fig.~\ref{fig:preparation}, these Pauli errors could be $[\sigma^y_1]$, $[\sigma^z_1]$, $[\sigma^x_2]$, $[\sigma^y_2]$, $[\sigma^z_2]$, $[\sigma^y_3]$ and $[\sigma^z_3]$, and all other errors are equivalent to these seven types of errors. 

Correlations may also occur in switchyards. To deal with these correlations, we can ensure that states from the same switchyard are utilised in different cluster-state qubits that are separated by distances much larger than the dimension of logical qubits. In this way, these correlations induced by switchyards never form a correlated error that can risk a logical qubit. 

The depolarising error reads
\begin{equation}
\mathcal{E} = (1-\epsilon)[\openone] + \frac{\epsilon}{3}([\sigma^x] + [\sigma^y] + [\sigma^z])
\end{equation}
Here, $\epsilon$ is the error rate, and the superoperator $[U]\rho = U\rho U^\dag$. 

\section{Numerical simulations}
\label{sec:AppNS}

\begin{figure*}[tbp]
\centering
\includegraphics[width=0.67\linewidth]{\figpath 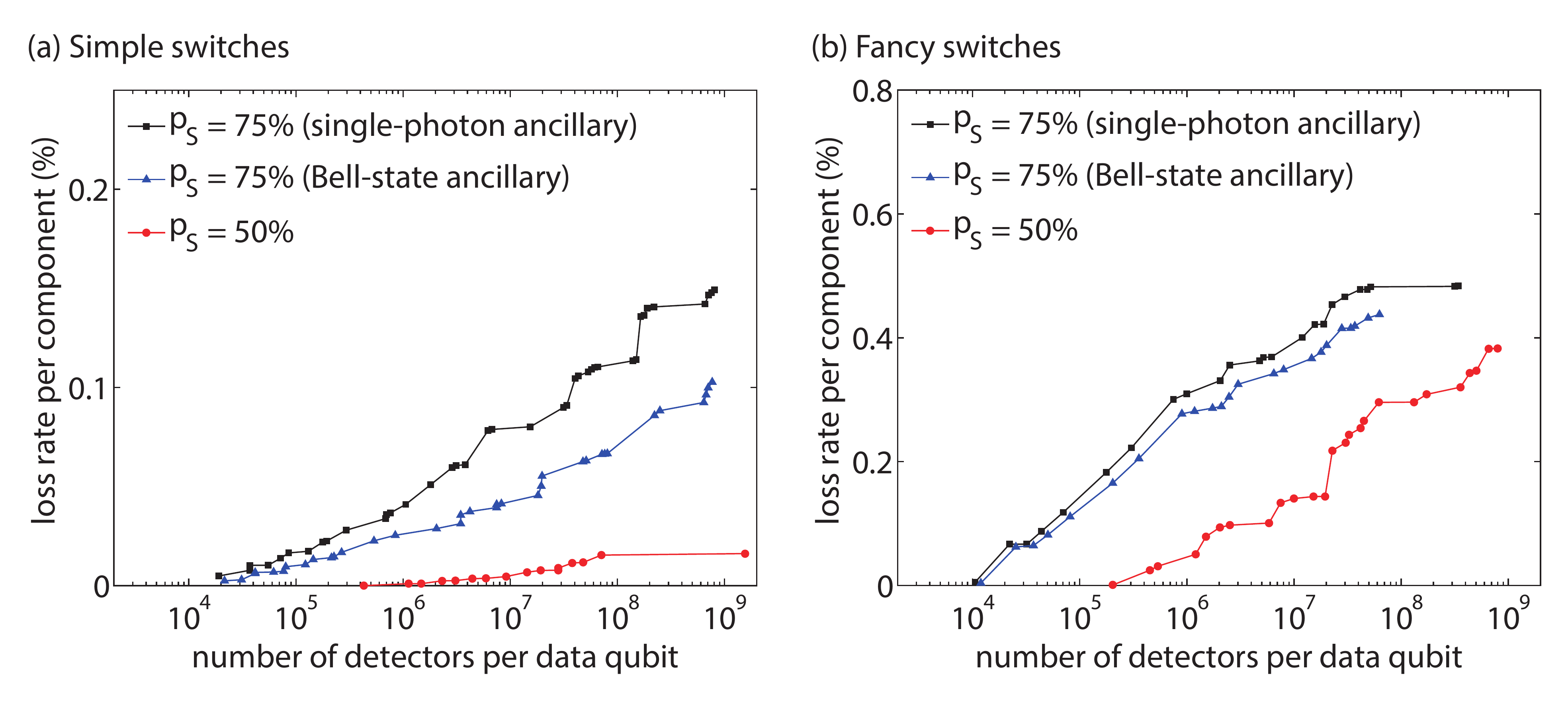}
\caption{
Thresholds of the loss rate per component $p$ as a function of the number of detectors per data qubit in the case with computational errors. All components have the same loss rate, i.e.~$\pro{e} = \pro{b} = \pro{d} = \pro{s} = \pro{m} = p$. And the rate of computational errors per component is $\eps{b} = \eps{d} = \eps{s} = 3\times 10^{-6}$. 
}
\label{fig:lossrate-error}
\end{figure*}

The threshold of fault-tolerant quantum computing is determined by evaluating $p_L$ and $p_P$, which respectively are the loss rate and phase-error rate of a cluster-state qubit, for the given loss rate and error rate per component. In Ref.~\cite{Barrett2010}, the inferred critical threshold is almost a straight line in the (loss rate, computational error rate) parameter space. For the 3D cluster state, the phase-error rate threshold without loss is $2.93\%$~\cite{Raussendorf2006}, and the loss rate threshold without error is $24.9\%$. Therefore, the threshold of $(p_L,p_P)$ is estimated as
$$
\frac{p_L}{24.9\%} + \frac{p_P}{2.93\%} = 1.
$$

To obtain thresholds of the loss rate per component without computational error, we have considered 2-level trees and 3-level trees [see Fig.~\ref{fig:lossrate-trees}] with branching numbers not larger than $20$ as building-block states, which includes $8400$ different tree structures in total. In the case that switchyards are composed by 2-input-2-output switches, we have considered configurations of the switch network for which the number of outputs for all of the switchyards is the same. We simulate output numbers $N = 2^n$ with $n = 0,1,\ldots,10$. For switchyards at the outputs of GHZ state factories, the input number is $M = m\times N$ with $m = 32,36,\ldots,256$ (the success rate of generating GHZ states is $1/32$~\cite{Varnava2008}). For switchyards for selecting successful Bell measurements, the input number is determined by the input number of GHZ-state switchyards, which is $\ceil(M/(32p_S))$, where $p_S = 50\%,75\%,87.5\%$ is the success rate of Bell measurements without photon loss. Similarly, for switchyards for selecting successfully generated Bell states, the input number is $\ceil(M/4)$, where we have used the circuit for generating Bell states with the success rate $1/8$, which can be boosted to $3/16$ if a switch is introduced~\cite{Browne2005}. Therefore, we have in total considered $627$ different configurations of the switch network composed by 2-input-2-output switches. In the case that each switchyard is a fancy switch with arbitrarily large input and output numbers, we have assumed that the ratio `output~number{\slash}input~number' equals the actual success rate (including the effect of photon loss) of corresponding operations. 

\begin{figure*}[tbp]
\centering
\includegraphics[width=0.9\linewidth]{\figpath 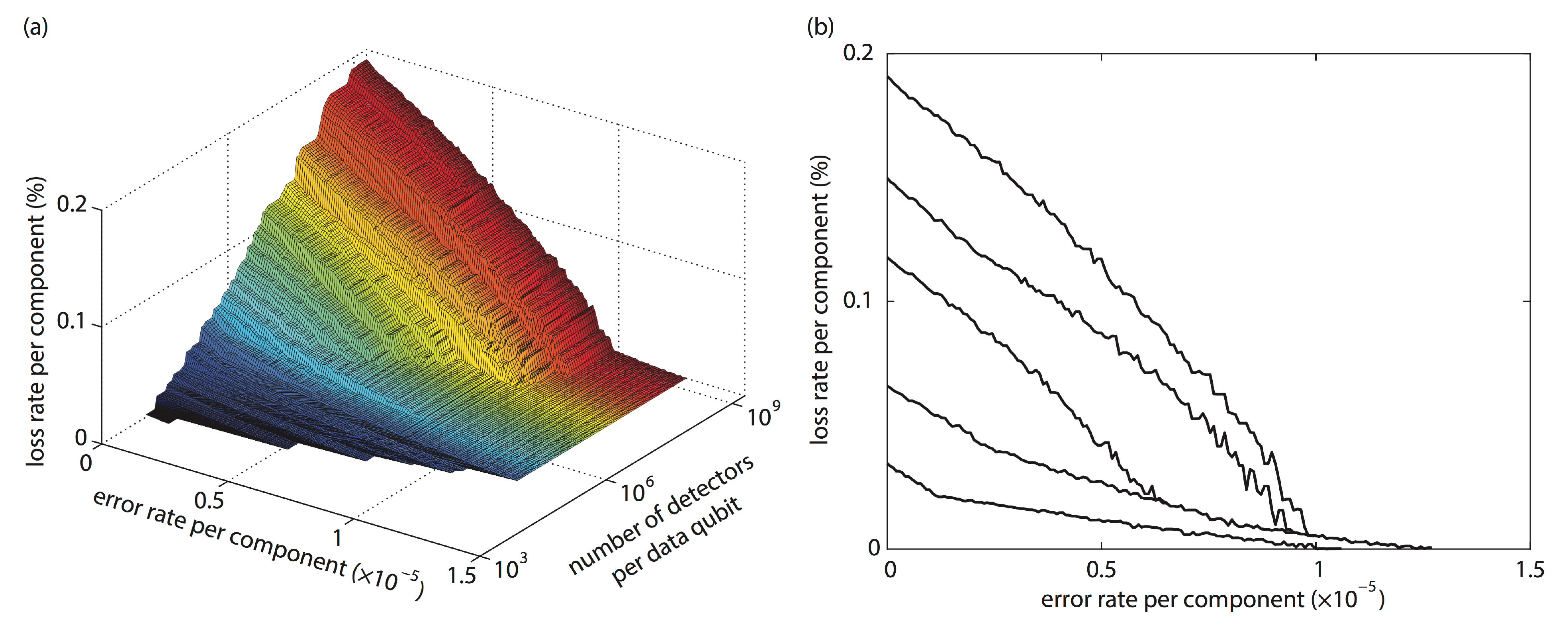}
\caption{
(a) The full data of Fig.~\ref{fig:3D}. (b) Thresholds on the loss rate per component $p$ for given the error rate per component $\epsilon$ with $10^5$, $10^6$, \ldots, $10^9$ detectors per data qubit from bottom to top. Fluctuations are due to the uncertainty of the Monte Carlo method. 
}
\label{fig:error}
\end{figure*}

For each curve in Fig.~\ref{fig:lossrate}~(a)~and~(c)-(f), thresholds of the loss rate per component are evaluated for $8400\times 627$ different protocols. Each protocol includes the building-block structure and the configuration of the switch network. Each curve is obtained as the envelope of these thresholds. For each curve in Fig.~\ref{fig:lossrate}~(b), thresholds of the loss rate per component are evaluated for $8400$ different protocols, which are only determined by building-block structures. 

To obtain thresholds of the loss rate per component with computational errors, we have selected about $500$ protocols from protocols that require not more than $2\times 10^9$ detectors for each case. These selected protocols are all close to the envelope, i.e. have the best performance of tolerating photon loss. Specifically, we have drawn a straight line connecting the highest point (corresponding to the protocol tolerates the highest loss rate per component) and the lowest point (corresponding to the protocol with the smallest number of detectors) on the envelope. This line is then shifted downwards until there are about $500$ protocols whose thresholds of the loss rate per component are above it. Thresholds of the loss rate per component with computational errors in Fig.~\ref{fig:3D} are obtained from these selected protocols. Computational errors are evaluated using Monte Carlo methods. In each protocol, for each value of the loss rate and the error rate, the phase-error rate on a cluster-state qubit is obtained with 100000 samples. 

\section{Bell measurements with entangled ancillary states}
\label{sec:AppBM}

In addition to the Bell measurement assisted by a Bell state (see Figs.~\ref{fig:lossrate}), which has the success probability $75\%$, we also have considered the Bell measurement assisted by a $4$-qubit GHZ state (see Fig.~\ref{fig:lossrate-875}), which has the success probability $87.5\%$~\cite{Grice2011}. The $4$-qubit GHZ state is prepared with two $3$-qubit GHZ states generated with the circuit in Fig.~\ref{fig:circuits}~(b). By using a PP operation, in which the Bell measurement is assisted by a Bell state, i.e.~the success probability is $75\%$, two $3$-qubit GHZ states can be fused into a $4$-qubit GHZ state. We find that further boosting the success probability of Bell measurements with more entangled ancillary photons is not helpful. 


\end{document}